\pdfoutput=1
\documentclass[%
 reprint,
 amsmath,
 amssymb,
 prl
]{revtex4-1}

\usepackage[dvipdfmx,dvipdfmx]{graphicx}
\usepackage{graphicx}
\usepackage{color}
\usepackage{siunitx}

\usepackage{subfigure}

\usepackage{epsfig}
\usepackage{float}
\usepackage{epstopdf}
\usepackage{amsmath}
\usepackage{amsfonts}
\newcommand{\abs}[1]{\left| #1 \right|}
\newcommand{\ket}[1]{\left | #1 \right \rangle}
\def\k(#1){|#1\rangle}

\newcommand{\tr}{{\rm \, Tr }\, }

\newcommand{\beq}{\begin{equation}}
\newcommand{\eeq}{\end{equation}}
\newcommand{\beqa}{\begin{eqnarray}}
\newcommand{\eeqa}{\end{eqnarray}}
\newcommand{\beqan}{\begin{eqnarray*}}
\newcommand{\eeqan}{\end{eqnarray*}}

\newcommand{\affA}{%
\affiliation{
 Center for Macroscopic Quantum States (bigQ), Department of Physics, Technical University of Denmark, Building 307, Fysikvej, 2800 Kgs.~Lyngby, Denmark}
     }

\begin{document}

\title{
%Tomography of a displacement photon counter with feedback operations
% Tomography of an adaptive measurement
% Tomography of a dynamically updated measurement
% Tomography of a measurement with feedback
% Tomography of a feedback measurement with photon counting
% Tomography of a qubit discriminator based on photon counting and real-time feedback 
Tomography of a feedback measurement with photon detection
}

%\date{\today}

\author{Shuro Izumi}
\affA
\author{Jonas S. Neergaard-Nielsen}
\affA
\author{Ulrik L. Andersen}
\affA

\begin{abstract}
% The projective measurement of a qubit in its computational basis as well as its conjugate basis is an incredibly important tool 
% in both fundamental studies and practical applications of quantum information science. For some qubits (e.g. spin and polarization qubits), these measurements are simple while for other qubits (e.g single-rail and cat qubits) it is notoriously difficult. 
% Here we experimentally demonstrate a measurement that is known to be, in principle, an ideal projection measurement onto the superposition of vacuum and single photon states -- the single-rail qubit.
% Our measurement consists of a displacement operation combined with a photon counting measurement followed by a real time feedback operation. We characterize the single-rail qubit projector by detector tomography and find that it can discriminate the conjugate basis states with a certainty of 96\%.
% Such a feedback controlled photon counter will facilitate the realization of quantum information protocols with single-rail qubits as well as the non-locality test of certain entangled states.

Quantum measurement is essential to both the foundations and practical applications of quantum information science.
Among many possible models of quantum measurement,
feedback measurements that dynamically update their physical structure are highly interesting due to their flexibility which enables a wide range of measurements that might otherwise be hard to implement.
Here we investigate by detector tomography a measurement consisting of a displacement operation combined with photon detection followed by a real time feedback operation.
We design the measurement in order to discriminate the superposition of vacuum and single photon states -- the single-rail qubit --  and find that it can discriminate the superposition states with a certainty of 96\%.
Such a feedback-controlled photon counter will facilitate the realization of quantum information protocols with single-rail qubits as well as the non-locality test of certain entangled states.
\end{abstract}

\maketitle

{\it Introduction.}---Measurement and discrimination of quantum states plays a fundamental role in quantum information processing \cite{Helstrom_book76_QDET, NielsenChuang,RevModPhys.79.135,GisinThew}. The information contained in qubits or higher-dimensional encodings can only be faithfully retrieved if high-fidelity readout schemes are available: The result of a quantum computation is obtained by projective measurements of the qubit states in the computational basis \cite{RevModPhys.79.135}, while the message or key sent over a communication channel is extracted by measurements that can efficiently discriminate some of the quantum states used as code words \cite{GisinThew,RevModPhys.74.145}.
% In many cases, it is obvious how to perform such measurements-indeed, physical implementations of qubits are not very useful unless they provide for a low-error measurement in the computational basis. 
% In other protocols or systems, it may not be easy or at all possible to implement a direct measurement that reaches the required performance. There, more advanced but indirect measurement schemes may be needed.
In some cases, it may not be easy or even possible to implement a direct measurement that attains the required performance. Therefore, more advanced but indirect measurement schemes may be needed.
As an example of this, relevant to coherent optical communication, two weak coherent states cannot be optimally distinguished using standard optical measurements like homodyne, heterodyne or photon detection \cite{Helstrom_book76_QDET,PhysRevA.78.022320}. However, a measurement strategy consisting of photon detection preceded by a displacement operation that is dynamically updated based on the photon detector outcomes is known to attain the optimal performance possible
% (the Helstrom bound) 
in the limit of infinitely fast feedback \cite{Dolinar73,CookMartinGeremia2007_Nature}. As a highly flexible measurement model, such a feedback measurement with its dynamically updated physical structure adapted to partial measurement outcomes may have a wide range of applications \cite{Bondurant93,izumi2012,izumi2013,Becerra13,Becerra15}. 
Indeed, it has also been shown to enable an arbitrary two-dimensional projection measurement \cite{TakeokaSasakiLutkenhaus2005,TakeokaSasakiLutkenhaus2006_PRL_BinaryProjMmt}.

A scenario where this measurement strategy will be highly beneficial is the discrimination of the two orthogonal superpositions of the vacuum and single photon state, $|\pm\rangle=\frac{1}{\sqrt{2}}(|0\rangle\pm |1\rangle)$. 
These are the conjugate basis states of the optical single-rail qubit where information is encoded in the photon number of a single optical mode \cite{LundRalph,morin2014remote,Jeong,PhysRevLett.118.160501,2019arXiv190508562D}.
% on single-rail qubit, e.g. https://journals.aps.org/pra/pdf/10.1103/PhysRevA.66.032307, https://journals.aps.org/prl/pdf/10.1103/PhysRevLett.118.160501, https://arxiv.org/pdf/1905.08562.pdf, https://iqst.ca/quantech/pubs/2006/inerconvertibility-ol.pdf]. 
This particular qubit encoding is interesting due to its natural relation to e.g. atomic and mechanical qubits \cite{Kurizki3866} and its convertibility with cat state qubits and polarization qubits \cite{morin2014remote,Jeong,PhysRevLett.118.160501,2019arXiv190508562D}. The computational basis states can be distinguished simply by a high-efficiency photodetector, but the states of the conjugate basis are not associated with simple physical observables and can therefore not be directly measured.
% Whereas other types of qubits usually have simple ways of converting between the $X$ and $Z$ bases, thereby making $X$ basis measurements possible, this is not true for the single-rail qubits—such a Hadamard transformation would require strong nonlinearities that could be hard to achieve \cite{LundRalph,Heaney}. 
One discrimination strategy for the superposition states is to apply a Hadamard transform to the state (converting $|\pm\rangle$ into $|0\rangle$ or $|1\rangle$) prior to the measurement of its photon number, but such a transformation is highly non-trivial, requiring a strong non-linearity \cite{LundRalph,Heaney}.
The task of making a projective measurement onto the conjugate basis for the single-rail qubit is therefore an obvious use case for the more feasible feedback measurement.
% The problem of making a projective measurement in the single-rail qubit $X$ basis is therefore an obvious use case for the more feasible feedback measurement.

We experimentally demonstrate the implementation of a feedback measurement at telecom wavelength with parameters optimized for a projective measurement onto the vacuum and single photon superposition states $|\pm\rangle$. With different parameter settings, the measurement could be adapted to arbitrary projections on the single-rail qubit Bloch sphere. The measurement consists of a displacement operation, photon detection and feedback for updating the displacement amplitude. We characterize the measurement by quantum detector tomography which only requires a collection of well-calibrated coherent states. Quantum detector tomography has been demonstrated for various types of static measurements \cite{Lundeen09, Brida,Zhang12,natarajan2013quantum,catprojection,Kennedytomography}, but not before for a dynamically updated measurement. As a figure of merit for the performance of the measurement, we evaluate the discrimination error for the two superposition states.

\begin{figure}[t]
\centering 
{
\includegraphics[width=1.0\linewidth]
{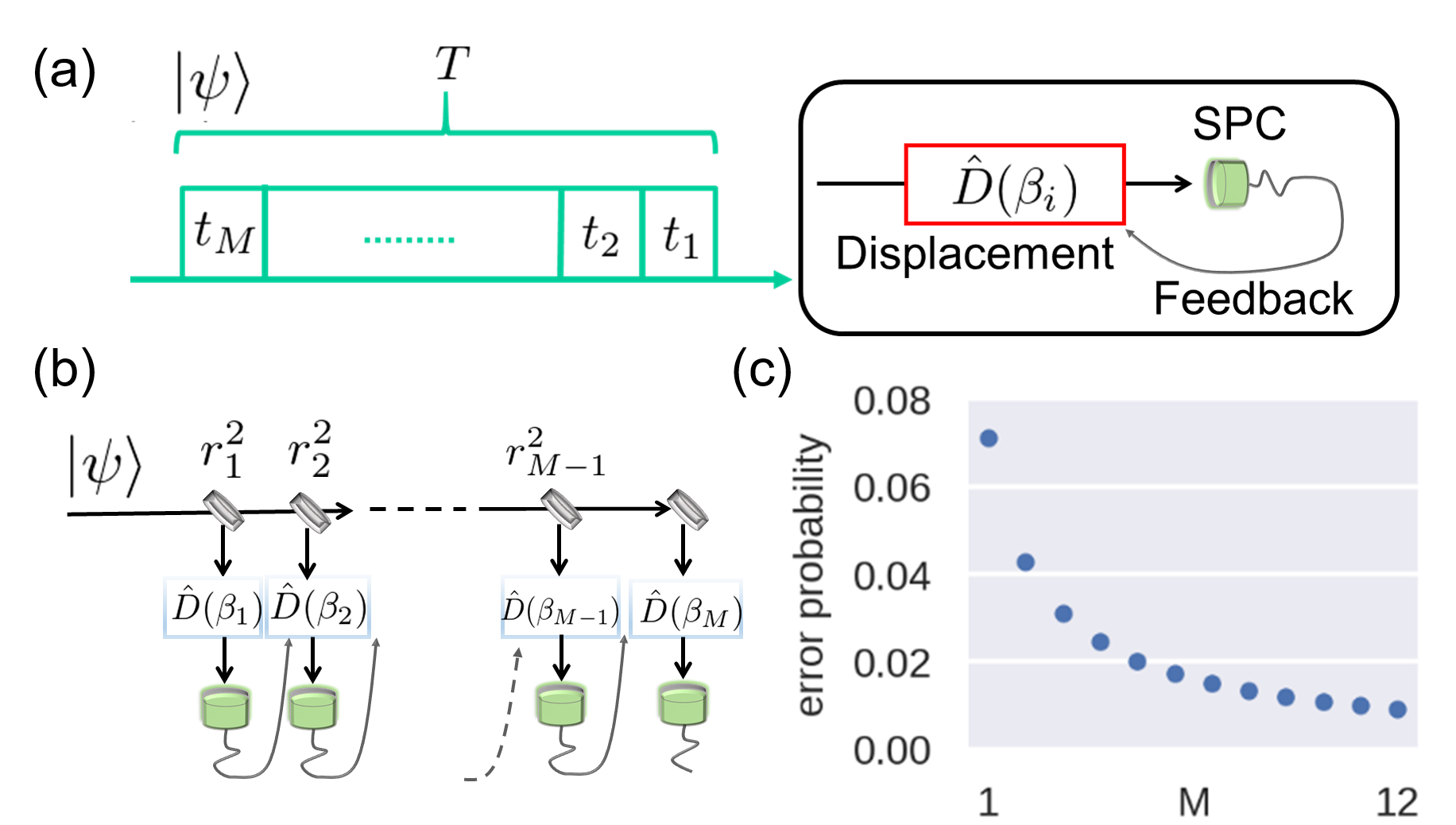}
}
\caption{
Schematic of the displacement followed by single photon counter (SPC) with feedback operations,
(a) temporal mode version
and
(b) spatial mode version.
(c) Performance of the displacement and SPC with finite feedback operations.
% $M=1$ is the displacement photon counting measurement without the feedback operation.
\label{Dolinar_theory}
}
\end{figure}

{\it Concept.}---The original proposal for quantum state discrimination using a displacement based photon counter with feedback operations shown in Fig.~\ref{Dolinar_theory}(a) consists of a displacement operation, a single photon counter (SPC) and real-time feedback of the SPC's measurement outcome to the displacement amplitude \cite{Dolinar73}.
An incoming quantum state with full time width $T$ is virtually divided into $M$ temporal mode bins with time widths $t_i$ for the $i$'th bin. The phase and amplitude of the displacement in each bin, $\beta_i$, is dependent on the photon counting history of the earlier time bins.
The measurement strategy can be equivalently analyzed using spatial modes, where the measurement consists of beam splitters (BSs) having reflectances $r_i^2$  and a displacement operation with the amplitude $\beta_i$ and a SPC in each of the $M$ spatial modes \cite{TakeokaSasakiLutkenhaus2005,TakeokaSasakiLutkenhaus2006_PRL_BinaryProjMmt}, as depicted in Fig.~\ref{Dolinar_theory}(b).

%%%%%%%%%%%%%%%%%%%%%%%%%%%%%%%%%%%%%%%%%%%%%%%%%%%%%%%%%%%%%%%%%%%%%%%%%%%%
%%%%%%%%%%%%%%%%%%%%%%%%%%%%%%%%%%%%%%%%%%%%%%%%%%%%%%%%%%%%%%%%%%%%%%%%%%%%%

For concreteness, we will describe the protocol in the setting of discrimination of the states $\ket{\pm}$.
The quantum state to be distinguished is split by the BSs and the first displacement operation is implemented such that the $\ket{+}$ state is displaced close to the vacuum state.
The displaced state is detected by the SPC, whose binary outcomes indicate whether the state is more likely to be $\ket{+}$ (off) or $\ket{-}$ (on).
Once an outcome from the SPC is obtained, one can calculate an {\it a posteriori} probability $P(\ket{\pm}|\{e_i\})$ for given outcomes $\{e_i\}$, where $e_i\in\{\mathrm{off},\mathrm{on}\}$.
The displacement amplitude is controlled dependent on the {\it a posteriori} probability, i.e., the most probable state is displaced close to the vacuum state.
By repeating the operations and detections recursively,
we conclude whether the state is $\ket{+}$ or $\ket{-}$ depending on the {\it a posteriori} probability.
The optimal strategy, it turns out, is to change the sign of the $i+1$'th displacement with respect to the $i$'th displacement if the $i$'th counter detects a photon and to maintain the phase otherwise.  
The conclusion of the state discrimination is then $\ket{+}$ if the total number of ``on'' events is even and $\ket{-}$ if it is odd \cite{TakeokaSasakiLutkenhaus2005,TakeokaSasakiLutkenhaus2006_PRL_BinaryProjMmt}.
The discrimination error approaches zero if $M\rightarrow\infty$.
We adopt the spatial mode analysis to investigate the performance of the measurement in the finite number case.

%%%%%%%%%%%%%%%%%%%%%%%%%%%%%%%%%%%%%%%%%%%%%%%%%%%%%%%%%%%%%%%%
%%%%%%%%%%%%%%%%%%%%%%%%%%%%%%%%%%%%%%%%%%%%%%%%%%%%%%%%%%%%%%%%
\begin{figure*}[t]
\centering
{
\includegraphics[width=1.00 \linewidth]
{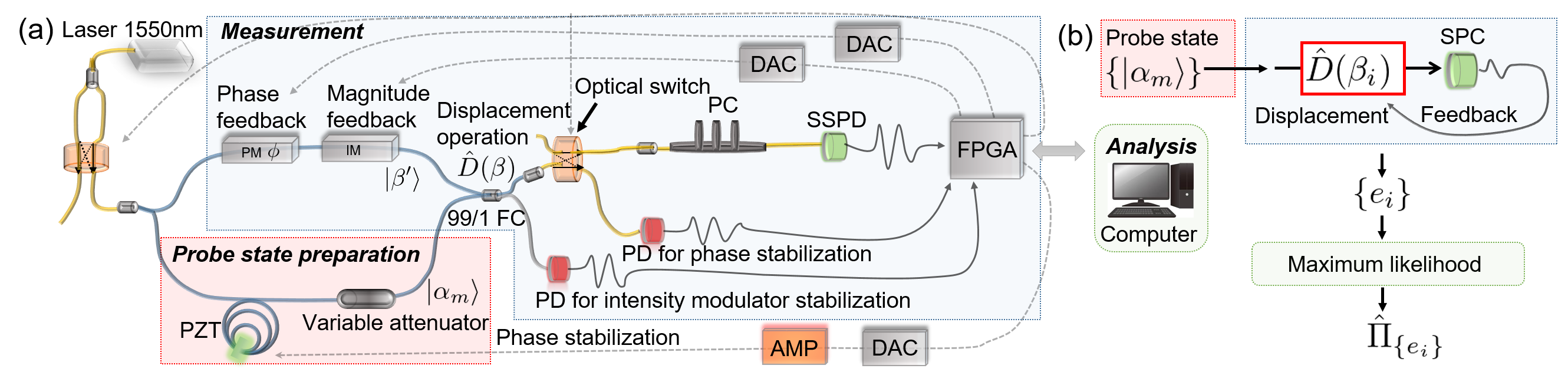}
}
\caption{
(a) Experimental setup. Blue and yellow fibers respectively represent polarization maintaining fiber and single mode fiber.
FC: fiber coupler,
PM: phase modulator,
IM: intensity modulator,
PZT: piezo transducer,
PC: polarization controller,
PD: photo detector,
SSPD: superconducting nanowire single photon detector. 
DAC: digital to analog converter,
AMP: amplifier.
(b) Schematic of the detector tomography.
% A maximum likelihood reconstruction (ML) is performed using the outcome statistics and the characteristics of the probe states to reconstruct the POVM $\hat{\Pi}_{\{e_i\}}$ for outcome $\{e_i\}$. 
}
\label{setup}
\end{figure*}
%%%%%%%%%%%%%%%%%%%%%%%%%%%%%%%%%%%%%%%%%%%%%%%%%%%%%%%%%%%%%%%%
%%%%%%%%%%%%%%%%%%%%%%%%%%%%%%%%%%%%%%%%%%%%%%%%%%%%%%%%%%%%%%%%

Each stage of the measurement, consisting of beam splitter, displacement, and SPC with the outcome $e_i$, can be considered as a single quantum operation $\mathcal{E}_{e_i}^{(i)}$ on the state that was output from the previous stage. The post-measurement state of the $i$'th stage is then $\hat\rho^{(i+1)} = \mathcal{E}_{e_i}^{(i)}(\hat\rho^{(i)}) / \tr \mathcal{E}_{e_i}^{(i)}(\hat\rho^{(i)})$, where the normalization factor is the probability of getting the outcome $e_i$. 
For an ``off'' detection, the map of the operation is $\mathcal{E}_\mathrm{off}^{(i)} (\hat\rho^{(i)}) = \hat{K}_0^{(i)} \hat\rho^{(i)} \hat{K}_0^{(i)\dagger}$, with the Kraus operator corresponding to zero photons at the detector given by 
$\hat{K}_0^{(i)} = e^{-\frac{1}{2} |\beta_i|^2} e^{-\hat{a} \beta_i^\ast r_i/t_i} e^{\hat{a}^\dagger \hat{a} \ln t_i}$ with beam splitter transmittance $t_i^2=1-r_i^2$ \cite{TakeokaSasakiLutkenhaus2005}.
Since the SPC cannot distinguish between one and more photons, the corresponding map for an ``on'' detection is $\mathcal{E}_\mathrm{on}^{(i)} (\hat\rho^{(i)}) = \sum_{n=1}^\infty \hat{K}_n^{(i)} \hat\rho^{(i)} \hat{K}_n^{(i)\dagger}$ with the $n$-photon Kraus operators $\hat{K}_n^{(i)} = \frac{1}{\sqrt{n!}} (\beta_i + \hat{a} \frac{r_i}{t_i})^n \hat{K}_0^{(i)}$.
The total probability for obtaining the series of detection events $\{e_1, \dots, e_M\}$ given an input state $\hat\rho$ is then the trace of the composition of the maps for each stage, $P(\{e_i\}|\hat\rho) = \tr \mathcal{E}_{e_M}^{(M)} \circ \cdots \circ \mathcal{E}_{e_1}^{(1)} (\hat\rho)$.
Equivalently, the probability can be written in terms of a POVM corresponding to that specific measurement outcome, $P(\{e_i\}|\hat\rho) = \tr [\hat\rho \hat\Pi_{e_1, \dots, e_M}]$. While the former formulation is most natural for understanding and modelling the iterative detection scheme, the latter is more relevant for the process of detector tomography which returns the elements of the POVM.
Thus, denoting the sets of all possible outcomes with even (odd) number of ``on'' events as $E$ ($O$), the error probability for the discrimination of the superposition states is given by,
%%%%%%%%%%%%%%%%%%%%%%%%%%%%%%%%%%%%%%%%%%%%%%%%%%%%%%%%%%%%%%%%%%%%%%%%%%%%%%%
% \begin{eqnarray}
% P_e = 
% \frac{1}{2}
% \Bigl(
% \hspace{-0.3cm}
% \sum_{\substack{\{e_i\}\in\\ \{\{e_i\}_{\mathrm{even}}\}}}
% \hspace{-0.3cm}
% P( \{e_i\}|\ket{-})
% +
% \hspace{-0.3cm}
% \sum_{\substack{\{e_i\}\in\\ \{\{e_i\}_{\mathrm{odd}}\}}}
% \hspace{-0.3cm}
% P( \{e_i\}|\ket{+})\Bigr).
% \label{eq2}
% \end{eqnarray}
\begin{eqnarray}
P_e = 
\frac{1}{2}
\Bigl(
\sum_{\{e_i\}\in E}
P( \{e_i\}|\ket{-})
+
\hspace{-0.3cm}
\sum_{\{e_i\}\in O}
\hspace{-0.3cm}
P( \{e_i\}|\ket{+})\Bigr).
\label{eq2}
\end{eqnarray}
%%%%%%%%%%%%%%%%%%%%%%%%%%%%%%%%%%%%%%%%%%%%%%%%%%%%%%%%%%%%%%%%%%%%%%%%%%%%%%%%

To illustrate the scheme and the optimizations involved, we first consider the feedback measurement of $\ket{\pm}$ in the case of $M=2$. Using the decision strategy and the expressions outlined above, we obtain the error probability for the feedback measurement with $M=2$ as,
%%%%%%%%%%%%%%%%%%%%%%%%%%%
\begin{eqnarray}
P_e^{M=2}&=& 
% \frac{1}{2} \left ( P(\{\text{off,off}\}|\ket{-}) + P(\{\text{on,on}\}|\ket{-}) + P(\{\text{off,on}\}|\ket{+}) + P(\{\text{on,off}\}|\ket{+}) \right) \\
% &=& \frac{P_\text{off,off}(\ket{-}) + P_\text{on,on}(\ket{-}) + P_\text{off,on}(\ket{+}) + P_\text{on,off}(\ket{+})} {2} \\
% &=&tyt\\
\frac{1}{2} - r\text{Re}[\beta_1] \text{e}^{-\abs{\beta_1}^2} \big(1 - \text{e}^{-\abs{\beta_{2,\text{on}}}^2} - \text{e}^{-\abs{\beta_{2,\text{on}}}^2}\big) \nonumber \\
&& - \sqrt{1-r^2} \text{Re}[\beta_{2,\text{on}}] \text{e}^{-\abs{\beta_{2,\text{on}}}^2} \big(1 - \text{e}^{-\abs{\beta_1}^2}\big) \nonumber \\
&& + \sqrt{1-r^2} \text{Re}[\beta_{2,\text{off}}] \text{e}^{-\abs{\beta_{2,\text{off}}}^2} \text{e}^{-\abs{\beta_1}^2} .
% \\
% &=&
% \frac{1}{2}-\mathrm{e}^{-\abs{\beta_1}^2}r\mathrm{Re}[\beta_1]-\mathrm{e}^{-\abs{\beta_{2,\mathrm{on}}}^2}\sqrt{1-r^2}\mathrm{Re}[\beta_{2,\mathrm{on}}]
% \nonumber
% \\
% &+&
% \mathrm{e}^{-\abs{\beta_1}^2-\abs{\beta_{2,\mathrm{off}}}^2}(r\mathrm{Re}[\beta_1]+\sqrt{1-r^2}\mathrm{Re}[\beta_{2,\mathrm{off}}])
% \nonumber
% \\
% &+&
% \mathrm{e}^{-\abs{\beta_1}^2-\abs{\beta_{2,\mathrm{on}}}^2}(r\mathrm{Re}[\beta_1]+\sqrt{1-r^2}\mathrm{Re}[\beta_{2,\mathrm{on}}]).
% \label{eq5}
\end{eqnarray}
%%%%%%%%%%%%%%%%%%%%%%%%%
The ``off'' and ``on'' indices on $\beta_2$ indicate that the amplitude of the second displacement depends on the outcome in the first channel.
The minimum achievable error probability is obtained to be $P_e^{M=2} = 0.040$ at the optimized parameter values $r^2=0.336$, $\beta_{1}=-0.643$, $\beta_{2,\mathrm{off}}=-0.514$, $\beta_{2,\mathrm{on}}=0.390$.
For comparison, the displacement and SPC scheme without feedback operation (i.e.~$M=1$) obtains $P_e^{M=1}\approx 0.071$, while homodyne detection would be able to achieve an error probability of 0.101 \cite{Kennedytomography}. 

Finding the ultimate performance for a given $M$
requires optimization over $M-1$ parameters for the BS reflection coefficients and $2^M-1$ parameters for the displacement magnitudes $\abs{\beta_i}$. This problem 
becomes intractable for large $M$. 
In a simplification of the scheme, we may assume that only the displacement phases, i.e. the sign of the $\beta_i$'s should depend on the outcome history, whereas the displacement magnitudes will be kept fixed at each stage, e.g. $\abs{\beta_{2,\mathrm{off}}}=\abs{\beta_{2,\mathrm{on}}}$.
It turns out (see later) that there is only a small penalty to pay for the error probability in using this simplified scheme.
The results of the minimization are plotted in Fig.~\ref{Dolinar_theory}(c).

%%%%%%%%%%%%%%%%%%%%%%%%%%%%%%%%%%%%%%%%%%%%%%%%%%%%%%%%%%%%%%%%%%%%%%%%%%%%%%%%%%%%%%%%%%%%%%%%%%%%%%%%%%%%%%%%%%%%%%%%%%%%%%%%%%%%%%%%%%%%%%%%%%%%%%%%%%%%%%%%%%%%%%%%%%%%%%%%%%%%%%%%%%%%%%%%%%%%%%%%%%%%%%%%%%%%%%%%%%%%%%%%%%%%%%
% \section{Experiment}\label{Sect:3}
%%%%%%%%%%%%%%%%%%%%%%%%%%%%%%%%%%%%%%%%%%%%%%%%%%%%%%%%%%%%%%%%%%%%%%%%%%%%%%%%%%%%%%%%%%%%%%%%%%%%%%%%%%%%%%%%%%%%%%%%%%%%%%%%%%%%%%%%%%%%%%%%%%%%%%%%%%%%%%%%%%%%%%%%%%%%%%%%%%%%%%%%%%%%%%%%%%%%%%%%%%%%%%%%%%%%%%%%%%%%%%%%%%%%%%

{\it Experiment.}---Fig.~\ref{setup}(a) illustrates our experimental setup.
Quantum detector tomography requires well characterized probe states that cover the Hilbert space of interest.
We use densely spaced coherent states as probes, since they are readily available and tomographically complete  \cite{Lundeen09}.
A continuous-wave, fiber-coupled laser at 1550 nm is split in two paths, one for preparation of the probe states and one for the reference field for the displacement operation.
The laser intensity is switched between high and low for the purposes of phase calibration/stabilization and measurement, respectively.
The intensity and the phase of the probe states are adjusted by a variable attenuator and a phase shifter that consists of a piezo transducer embedded in a circular mount with an optical fiber looped around.
We prepare probe states with 4 weak magnitudes $\abs{\alpha}\approx\{0.4, 0.6, 0.8, 1.0 \}$ and 8 phase conditions $j\pi/4,\ j=0,\ldots,7$ as well as the vacuum state to characterize our measurement in the two-dimensional Hilbert space spanned by $\ket{0}$ and $\ket{1}$.
The displacement operation is physically implemented with a 99:1 fiber coupler.
Its magnitude and direction is controlled by a phase and an intensity modulator.
When the laser intensity is high (locking mode), a switch directs the displaced probe to a conventional photo detector for stabilization of the relative phase between probe and reference.
When the intensity is low (measurement mode), the displaced probe state is detected by a superconducting nanowire single photon detector (SSPD) \cite{SSPD,SSPD2}.
A field programmable gate array (FPGA) counts the electrical signal from the SSPD and rapidly changes---dependent on the measurement outcome---the voltages applied to the intensity and phase modulators. 
The procedure is repeated 10,000 times for each of the probe states.
In Fig.~\ref{setup}(b), we show a simple schematic of detector tomography with coherent states.
The POVM elements $\hat{\Pi}_{\{e_i\}}$ corresponding to the outcome $\{e_i\}$ of the measurement are reconstructed following the maximum likelihood procedure using the known density matrices of the probe states and the distribution of the outcomes \cite{Fiurasek}.

Losses in the switch and other optical components and the finite detection efficiency of the SSPD limit the performance of our measurement.
The total transmittance from the 99:1 fiber coupler to the fiber right before the SSPD is measured to be $65\%$.
A benefit of the SSPD is that by changing the applied bias voltage, one can tune the trade-off between high detection efficiency and low dark count rate. We set the efficiency to $\sim$51\% which results in a dark count rate of $\sim$20 counts per second.
As a proof of concept and to highlight the functionality of the measurement, we choose to disregard the finite overall efficiency $\eta$. In practice, we do this by calibrating the magnitude of the probe states and displacements by the actual count rate of the SSPD. This corresponds to a rescaling of the probe coherent state amplitudes, $\sqrt{\eta}\alpha \rightarrow \alpha$.
We note that after completion of this work, we further improved the transmittance efficiency to $90\%$ and the visibility to $99.6\%$ while we also used another SSPD with up to $73\%$ detection efficiency \cite{izumi2020experimental}.
%%%%%%%%%%%%%%%%%%%%%%%%%%%%%%%%%%%%%%%%%%%%%%%%%%%%%%%%%%%%%%%%%%%%%%%%%%%
\begin{figure}[h!]
\centering
{
\includegraphics[width=1\linewidth]
{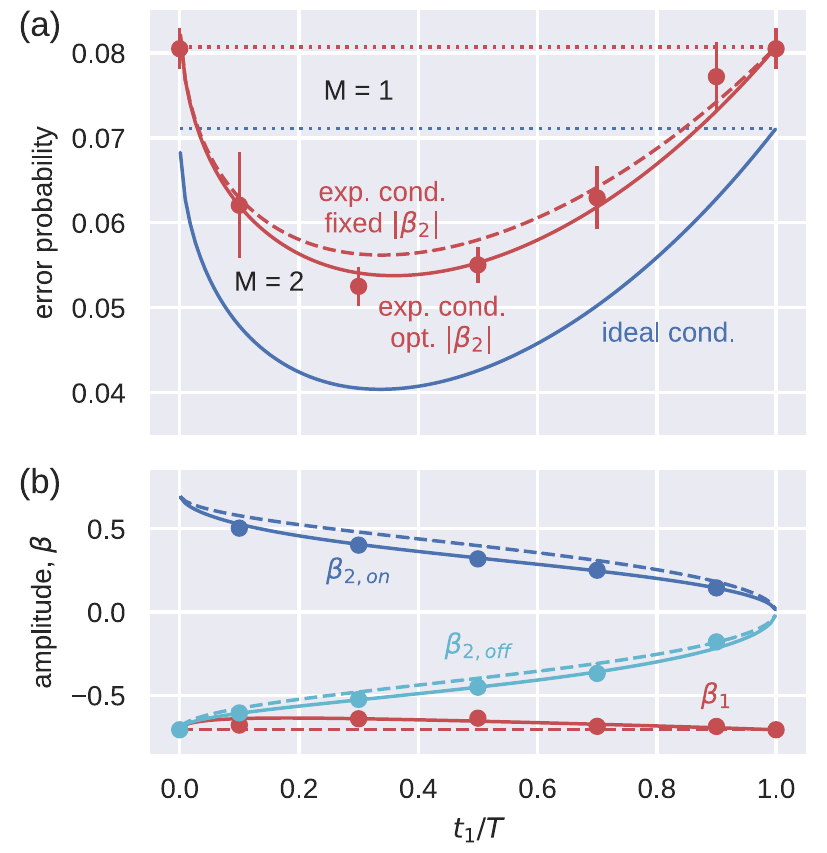}
}
\caption{
(a)
Error probability for the discrimination of the $\ket{\pm}$ states using our displacement and SPC with a single feedback operation ($M=2$) as a function of the relative temporal width of the first time bin, $t_1/T$.
(b) Amplitude conditions for the displacement operation.  
$\beta_1$ is the displacement amplitude in the first time bin, while $\beta_{2,\mathrm{off}}$ and $\beta_{2,\mathrm{on}}$ are for the second time bin if the outcome of the first time bin is ``off'' or ``on''. The curves indicate the theoretical optimum values as a function of $t_1$. Solid curves are for second-stage magnitudes adapted to the first stage's detector outcome, while dashed curves are for fixed magnitudes ($|\beta_{2,\text{off}}| = |\beta_{2,\text{on}}|$).
\label{Dolinar_M=2}
}
\end{figure}
%%%%%%%%%%%%%%%%%%%%%%%%%%%%%%%%%%%%%%%%%%%%%%%%%%%%%%%%%%%%%%%%%%%%%%%%%%%%

We first explore the $M=2$ case in detail, investigating the error probability with variable beam-splitter ratio and optimized displacement amplitudes.
The quantum states of the probes are defined within a rectangular temporal mode of length $T$ = $100\,\mu$s.
The feedback bandwidth of our experiment is limited by a digital analog converter whose bandwidth is roughly 1~MHz.
The delay of the feedback operations degrades the discrimination error.
Therefore, we discard counts observed in a $2\,\mu$s time interval after the first time bin, which corresponds to $2\%$ loss.
The delay analysis is further discussed in Supplemental Material.

Experimental results of the estimated error probability for $M=2$ with various settings of $t_1$
(corresponding to the beam-splitter ratio in the spatial picture)
are shown in Fig.~\ref{Dolinar_M=2}(a).
Red data points are the experimentally estimated error probabilities.
The mean values and the error bars are evaluated from five independent procedures.
The blue and red solid curves represent, respectively, the expected performance of the feedback measurements in the ideal case and with experimental imperfections, the non-unit visibility $\xi=0.98$, the dark count noise $2.38\times10^{-3}$ counts/state and $2\%$ loss due to the feedback delay.
The slightly degraded performance shown by the red dashed curve is 
what would be obtained by keeping the displacement magnitude fixed
($\abs{\beta_{2,\mathrm{off}}}=\abs{\beta_{2,\mathrm{on}}}$).
Measurement without feedback operation with and without the experimental imperfections are shown by red and blue dot lines.
Figure \ref{Dolinar_M=2}(b) depicts the experimental conditions as well as the theoretical optimum values of the displacement amplitudes.
The experimental results agree well with the theoretical prediction and the optimization of the feedback timing is essential to achieve the minimum error probability for a given $M$.

%%%%%%%%%%%%%%%%%%%%%%%%%%%%%%%%%%%%%%%%%%%%%%%%%%%%%%%%%%%%%%%%
%%%%%%%%%%%%%%%%%%%%%%%%%%%%%%%%%%%%%%%%%%%%%%%%%%%%%%%%%%%%%%%%
\begin{figure}[t]
\centering
{
\includegraphics[width=1 \linewidth]
{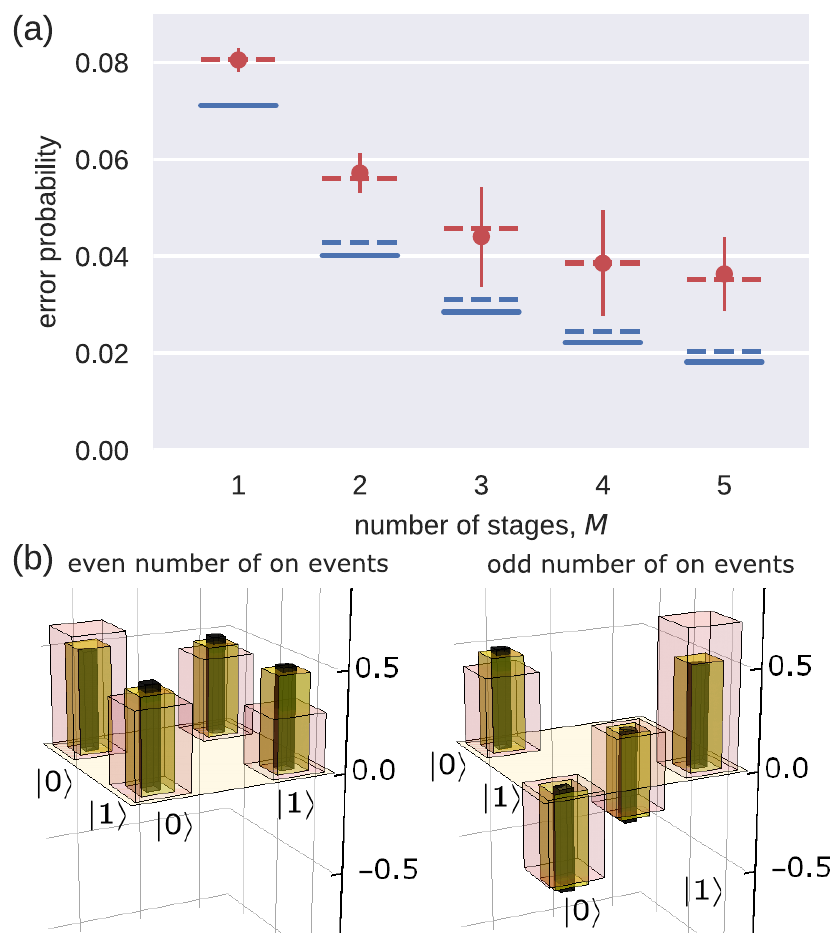}
}
\caption{
(a) 
Estimated error probabilities for the discrimination of $\ket{\pm}$ using various number of feedback operations. The points are experimentally obtained values with error bars. The red lines indicate the values expected from the model, while the blue lines indicate the ideal performance with no experimental imperfections. Dashed lines are for fixed displacement magnitudes at each stage, while the solid lines are for magnitudes adapted to the count history. 
% (b) Experimental conditions of the displacement magnitudes and time bin divisions. The thickness of the top lines indicate the uncertainty ($\pm$ 1 standard deviation) of the magnitudes, while the solid gray lines show the ideal values for minimum error discrimination. The gaps between time bins are the periods that are discounted to alleviate errors due to the finite feedback time.
(b) Reconstructed POVMs for even number (left) and odd number (right) of ``on'' events. Black, yellow and red bars represent POVM elements for the ideal single-rail projector, the experimental results for $M=5$ and $M=1$.
}
\label{Dolinar_M=5}
\end{figure}
%%%%%%%%%%%%%%%%%%%%%%%%%%%%%%%%%%%%%%%%%%%%%%%%%%%%%%%%%%%%%%%%
%%%%%%%%%%%%%%%%%%%%%%%%%%%%%%%%%%%%%%%%%%%%%%%%%%%%%%%%%%%%%%%%
Next, we investigate the performance of the measurement scheme for up to five feedback stages. 
The experimentally obtained error probabilities are plotted in Fig.~\ref{Dolinar_M=5}(a).
% and expected theoretical values are shown by red dashed lines.
The mean values and the error bars are calculated from 5 independent procedures for $M=1, 2$ and 10 independent procedures for $M=3, 4, 5$.
The corresponding displacement conditions realized in the experiment are shown in Supplemental Material.
% The displacement operation at each stage is set to a fixed magnitude for technical simplicity since the gain resulting from the counting history is small.
The delay time becomes dominant if the number of feedback operations is increased and the improvement of the error probability owing to the feedback operation could be saturated.
Therefore, we increase the time width for the probe state as $T=100 \times (M-1)\mu$s such that the delay loss can be constant 2$\%$ and decrease the bias voltage applied to the SSPD to get the constant dark counts $\sim2\times10^{-3}$ counts/state.
In Fig.~\ref{Dolinar_M=5}(b), we show the real part of the experimentally reconstructed POVMs of our measurement for $M=1$ and $M=5$ in the two-dimensional Hilbert space of interest.
% and compare with the ideal projector.
This makes it clear how feedback and adaptively updated parameters makes it possible to attain a near-perfect projective measurement.
Although the actual performance does not reach the ideal one (due to the experimental imperfections), we observe a clear improvement of the error probability by increasing the number of feedback operations.

{\it Summary.}---We experimentally demonstrated a measurement system designed for the discrimination of single-rail qubits. It is composed of a displacement operation, a single photon counter and feedback operations that depend on the outcome of the photon detections.
Our measurement was characterized by quantum detector tomography using coherent state probes.
We first investigated the error probability attainable by our measurement with a single feedback operation for varying timing conditions, showing the importance of optimizing this parameter. 
Secondly, we showed that the expected discrimination error can be improved by increasing the number of feedback operations.

We expect that our projector will pave the way for various applications in quantum information science utilizing the single-rail optical qubit. Moreover, the projector can be used to demonstrate quantum non-locality between many parties using a multi-mode delocalized single photon state (also known as a W state) \cite{Greenberger,WisemanJones,Heaney,ChavesBrask,Laghaout2},
which in turn will provide device-independent security in quantum communication \cite{Acin,Branciard}, and may lead to remote preparation of the superposition state \cite{Ozdemir,BabichevPRL,RalphLundWiseman,Pozza}.
% Furthermore, it is interesting to note that the displacement and SPC with {\it optimized} feedback timing can also provide an advantage for coherent state discrimination in practical scenarios with finite feedback bandwidth \cite{izumi2012}.

From a practical point of view,  
The ultimate speed of our feedback system would be limited by the dead time of the SSPD which is about 50 ns
while the current speed limitation due to the DAC can be solved by installing one with higher performance.
Further improvement of the feedback bandwidth could be obtained by a multi-pixel SSPD with effectively short dead time \cite{doi:10.1063/1.4921318}
or multiple SSPDs and a fast switch.
If one needs to perform the feedback measurement with very short laser pulses, the spatial configuration with delay lines (Fig.~\ref{Dolinar_theory}(b)) could be a practical direction to repeatedly implement the feedback.

\begin{acknowledgments}
We thank T. Yamashita, S. Miki, H. Terai for providing and installing the superconducting nanowire single photon detector.
This project was supported by Grant-in-Aid for JSPS Research Fellow, by VILLUM FONDEN via the Young Investigator Programme (Grant no. 10119) and by the Danish  National Research Foundation through the Center for Macroscopic Quantum States (bigQ DNRF142). 

\end{acknowledgments}

%\bibliography{reference}

\end{document}

\title{Supplemental material
}

%\date{\today}

\author{Shuro Izumi}
\affA
\author{Jonas S. Neergaard-Nielsen}
\affA
\author{Ulrik L. Andersen}
\affA

\maketitle

%%%%%%%%%%%%%%%%%%%%%%%%%%%%%%%%%%%%%%%%%%%%%%%%%%%%%%%%%%%%%%%%%%%%%%%%%%%%%%%%%%%%%%%%%%%%%%%%%%%%%
%%%%%%%%%%%%%%%%%%%%%%%%%%%%%%%%%%%%%%%%%%%%%%%%%%%%%%%%%%%%%%%%%%%%%%%%%%%%%%%%%%%%%%%%%%%%%%%%%%%%%
\section{Fundamental bound of the discrimination error with linear loss}\label{Sect:1}
%%%%%%%%%%%%%%%%%%%%%%%%%%%%%%%%%%%%%%%%%%%%%%%%%%%%%%%%%%%%%%%%%%%%%%%%%%%%%%%%%%%%%%%%%%%%%%%%%%%%%
%%%%%%%%%%%%%%%%%%%%%%%%%%%%%%%%%%%%%%%%%%%%%%%%%%%%%%%%%%%%%%%%%%%%%%%%%%%%%%%%%%%%%%%%%%%%%%%%%%%%%
We analyze the fundamental bound of the discrimination error for the superposition states with a linear loss $\eta$.
The loss can be discussed by a beam splitter model and the superposition states after the loss can be described as
%%%%%%%%%%%%%%%%%%%%%%%%%%%
\begin{eqnarray}
\hat{\rho}_{\pm}^{\eta} =
\begin{array}{l}
\frac{1}{2}\left[
\begin{array}{cc}
2-\eta& \pm\sqrt{\eta}\\
\pm\sqrt{\eta} & \eta \\
\end{array}
\right]
\end{array}
\label{eq1}
\end{eqnarray}
%%%%%%%%%%%%%%%%%%%%%%%%%%%
in the photon number basis.
Thus, the error probability for the superposition states with an equal prior probability is,
%%%%%%%%%%%%%%%%%%%%%%%%%%%
\begin{eqnarray}
P_e&=& \frac{1}{2}(\tr{[\hat{\rho}_{+}^{\eta}\hat{\Pi}_-]}+\tr{[\hat{\rho}_{-}^{\eta}\hat{\Pi}_+]})
\nonumber
\\
&=&
\frac{1}{2}(1-\tr{[(\hat{\rho}_{+}^{\eta}-\hat{\rho}_{-}^{\eta})\hat{\Pi}_+]}),
\label{eq2}
\end{eqnarray}
%%%%%%%%%%%%%%%%%%%%%%%%%%%
where $\hat{\Pi}_{\pm}$ are the POVM for concluding the input state as $\ket{\pm}$. 
The minimum error probability for the discrimination of $\hat{\rho}_{\pm}^{\eta}$ is obtained by maximizing 
$\tr{[(\hat{\rho}_{+}^{\eta}-\hat{\rho}_{-}^{\eta})\hat{\Pi}_+]}$ and given by,
%%%%%%%%%%%%%%%%%%%%%%%%%%%
\begin{eqnarray}
P_e&=& \frac{1}{2}(1-\lambda)
\nonumber
\\
&=&
\frac{1}{2}(1-\sqrt{\eta}),
\label{eq3}
\end{eqnarray}
%%%%%%%%%%%%%%%%%%%%%%%%%%%
where $\lambda=\sqrt{\eta}$ is a positive eigenvalue of the operator $(\hat{\rho}_{+}^{\eta}-\hat{\rho}_{-}^{\eta})$ \cite{Helstrom_book76_QDET}.
An optimal measurement to accomplish the error probability Eq.(\ref{eq3}) is given by a projector onto the superposition basis $\ket{\pm}$ irrespective of the loss.

In addition to the linear loss, various experimental imperfections cause the degradation of the performance and
we show a recipe to calculate the error probability for the measurement consisting of the displacement and the SPC with the experimental imperfections \cite{doi:10.1080/09500340903203103,6600857}.
In order to analyze the imperfection of the displacement operation due to limited visibility in the interference of the input and reference fields, we assume that the state is split into two modes 
where mode $0$ does not interfere with the coherent state for the displacement operation and mode $1$ is displaced with $\hat{D}_{1}(\sqrt{\xi}\beta)$. Both modes are detected by the SPC with imperfections. 
By denoting the visibility of the displacement operation as $\xi$,
the state is divided into two modes,
%%%%%%%%%%%%%%%%%%%%%%%%%%%
\begin{eqnarray}
&&\hat{B}(\xi)\ket{\pm}_1\ket{0}_2= 
\nonumber
\\
&&
\{\ket{0}_0\ket{0}_1+(\sqrt{\xi}\ket{0}_0\ket{1}_1\pm\sqrt{1-\xi}\ket{1}_0\ket{0}_1)\}/\sqrt{2}.
\label{eq4}
\end{eqnarray}
%%%%%%%%%%%%%%%%%%%%%%%%%%%
The probability of having ``off'' outcome is given by a product of the ``off'' probability for each mode.
Therefore, the POVM for the final outcomes is described as,
%%%%%%%%%%%%%%%%%%%%%%%%%%%%%%%%%%%%%%%%%%%%%%%%%%%%%%%%%%%%%%%%%%%%%%%%%%%%%%%
\begin{eqnarray}
\hat{\Pi}_{\mathrm{off}}&=&\mathrm{e}^{-\nu-(1-\xi)\eta\beta^2}\hat{\Pi}^{\mathrm{off}}_0\otimes \hat{D}_{1}^{\dagger}(\sqrt{\xi}\beta)\hat{\Pi}^{\mathrm{off}}_1\hat{D}_1(\sqrt{\xi}\beta),
\nonumber
\\
\hat{\Pi}_{\mathrm{on}}&=&\hat{I}-\hat{\Pi}_{\mathrm{off}},
\label{eq5}
\end{eqnarray}
%%%%%%%%%%%%%%%%%%%%%%%%%%%%%%%%%%%%%%%%%%%%%%%%%%%%%%%%%%%%%%%%%%%%%%%%%%%%%%%%
where $\hat{\Pi}^{\mathrm{off}}_i=\sum_{n=0}^{\infty}(1-\eta)^n\ket{n}\bra{n}$ is the probability of having ``off'' from the SPC for $i$ mode.
The dark count rate (defined as the probability per state of getting at least one dark count) and the detection efficiency of the SPC are represented by $\nu$ and $\eta$ respectively.
An additional unwanted count due to the coherent state for the displacement operation that does not interfere with the superposition state induces ``on'' event irrespective of the states to be discriminated.
Hence,
the error probability for the discrimination of the superposition states with the non-ideal measurement is, 
%%%%%%%%%%%%%%%%%%%%%%%%%%%%%%%%%%%%%%%%%%%%%%%%%%%%%%%%%%%%%%%%%%%%%%%%%%%%%%%
\begin{eqnarray}
P_e&=&
\frac{1}{2}( {}_{1}\bra{0}{}_{0}\bra{+}\hat{B}(\xi)^{\dagger}\hat{\Pi}_{\mathrm{on}}\hat{B}(\xi)\ket{+}_0\ket{0}_1
\nonumber
\\
&+&
{}_{1}\bra{0}{}_{0}\bra{-} \hat{B}(\xi)^{\dagger}\hat{\Pi}_{\mathrm{off}}\hat{B}(\xi)\ket{-}_0\ket{0}_1)
\nonumber
\\
&=&
\frac{1}{2}+\eta\xi\beta\mathrm{e}^{-\nu-\eta\beta^2}.
\label{eq6}
\end{eqnarray}
%%%%%%%%%%%%%%%%%%%%%%%%%%%%%%%%%%%%%%%%%%%%%%%%%%%%%%%%%%%%%%%%%%%%%%%%%%%%%%%%
Since derivation of the Kraus operator is not as straightforward as the ideal case, finding the analytical expression for the state evolution is not trivial.
Nevertheless, we can calculate the state evolution using the method outlined above and obtain the theoretical error probability for the feedback measurement with the imperfections up to 5 stages, which is shown in Fig.~4(a) in the main text. 

%%%%%%%%%%%%%%%%%%%%%%%%%%%%%%%%%%%%%%%%%%%%%%%%%%%%%%%%%%%%%%%%%%%%%%%%%%%%%%%%%%%%%%%%%%%%%%%%%%%%%
%%%%%%%%%%%%%%%%%%%%%%%%%%%%%%%%%%%%%%%%%%%%%%%%%%%%%%%%%%%%%%%%%%%%%%%%%%%%%%%%%%%%%%%%%%%%%%%%%%%%%
\section{Experimental procedure}\label{Sect:1}
%%%%%%%%%%%%%%%%%%%%%%%%%%%%%%%%%%%%%%%%%%%%%%%%%%%%%%%%%%%%%%%%%%%%%%%%%%%%%%%%%%%%%%%%%%%%%%%%%%%%%
%%%%%%%%%%%%%%%%%%%%%%%%%%%%%%%%%%%%%%%%%%%%%%%%%%%%%%%%%%%%%%%%%%%%%%%%%%%%%%%%%%%%%%%%%%%%%%%%%%%%%
%%%%%%%%%%%%%%%%%%%%%%%%%%%%%%%%%%%%%%%%%%%%%%%%%%%%%%%%%%%%%%%%%%%%%%
%%%%%%%%%%%%%%%%%%%%%%%%%%%%%%%%%%%%%%%%%%%%%%%%%%%%%%%%%%%%%%%%%%%%%%
\begin{figure}[t]
\centering
{
\includegraphics[width=1.0\linewidth]
{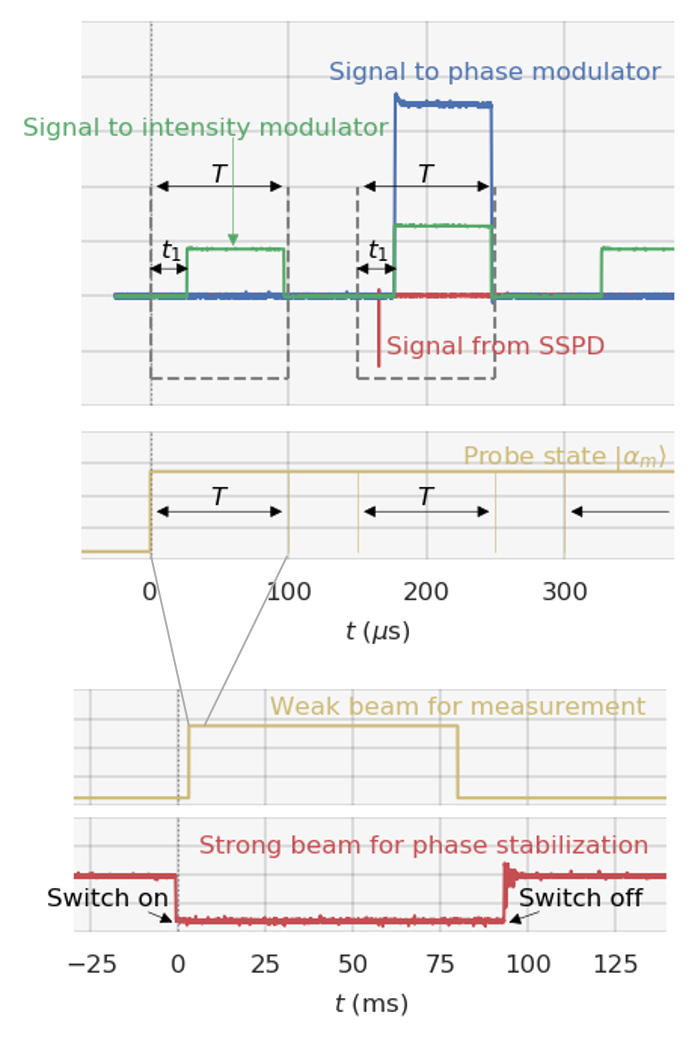}
}
\caption{
From bottom to top, an output from the interferometer detected by the photo detector, a weak signal injected into the interferometer, the probe state with the time length $T=100\mu$s, a signal from the SSPD and the feedback signals applied to the phase and intensity modulators. 
(Note that the displacement magnitude is decreased by increasing the bias voltage applied to the intensity modulator.) 
\label{Experiment_detail}
}
\end{figure}
%%%%%%%%%%%%%%%%%%%%%%%%%%%%%%%%%%%%%%%%%%%%%%%%%%%%%%%%%%%%%%%%%%%%%%
%%%%%%%%%%%%%%%%%%%%%%%%%%%%%%%%%%%%%%%%%%%%%%%%%%%%%%%%%%%%%%%%%%%%%%
Figure \ref{Experiment_detail} shows the procedure of our experiment for the feedback measurement with $M=2$.
In order to prepare the probe state with the targeted phase condition,
the interferometer is stabilized with a strong laser beam and a conventional photo detector.
The relative phase can be set to an arbitrary phase condition except close to $\theta=0, \pi$ by actively controlling a piezo transducer (PZT).
The probe state with the condition $\theta=0, \pi$ is prepared such that the phase is first stabilized to $\theta=\pi/2$ and an offset, that is calibrated to induce an additional $\pi/2$ shift, is added to the PZT after releasing the phase stabilization.
In the data acquisition period, we release the phase stabilization and switch the input to the interferometer to be a weak field.
The number of data points that can be acquired after releasing the phase stabilization is limited by the passive stability of the relative phase between the probe state and the displacement beam.
We acquire 500 data points in each data acquisition period, which takes $\sim 80$ ms including the switching time ($\sim 3$ ms) and repeat the procedure 20 times to get the total $10,000$ points for each probe state.
We set 50 $\mu$s time interval between each probe state in order to avoid phase drift due to the thermal effect of the phase and intensity modulators.
A program running on an FPGA detects the electrical signal from the SSPD and changes the voltages applied to the phase and the intensity modulators depending on the outcome from the SSPD for the feedback operation.

The output power from the intensity modulator slowly drifts with time since the modulator consists of a small interferometer. 
To implement reliable displacement operation with well characterized magnitude, while stabilizing the relative phase, we stabilize the intensity modulator by monitoring the output from the 99:1 fiber coupler by a photo detector and feedback-controlling the offset voltage applied to the intensity modulator.

%%%%%%%%%%%%%%%%%%%%%%%%%%%%%%%%%%%%%%%%%%%%%%%%%%%%%%%%%%%%%%%%%%%%%%%%%%%%%%%%%%%%%%%%%%%%%%%%%%%%%%%%%%%%%%%%%%%%%%%%%%%%%%%%%%%%%%%%%%%%%%%%%%%%%%%%%%%%%%%%%%%%%%%%%%%%%%%%%%%%%%%%%%%%%%%%%%%%%%%%%%
\section{Delay analysis of feedback operation}\label{Sect:2}
%%%%%%%%%%%%%%%%%%%%%%%%%%%%%%%%%%%%%%%%%%%%%%%%%%%%%%%%%%%%%%%%%%%%%%%%%%%%%%%%%%%%%%%%%%%%%%%%%%%%%%%%%%%%%%%%%%%%%%%%%%%%%%%%%%%%%%%%%%%%%%%%%%%%%%%%%%%%%%%%%%%%%%%%%%%%%%%%%%%%%%%%%%%%%%%%%%%%%%%%%%
%%%%%%%%%%%%%%%%%%%%%%%%%%%%%%%%%%%%%%%%%%%%%%%%%%%%%%%%%%%%%%%%%%%%%%
%%%%%%%%%%%%%%%%%%%%%%%%%%%%%%%%%%%%%%%%%%%%%%%%%%%%%%%%%%%%%%%%%%%%%%
\begin{figure}[b]
\centering
{
\includegraphics[width=1.0\linewidth]
{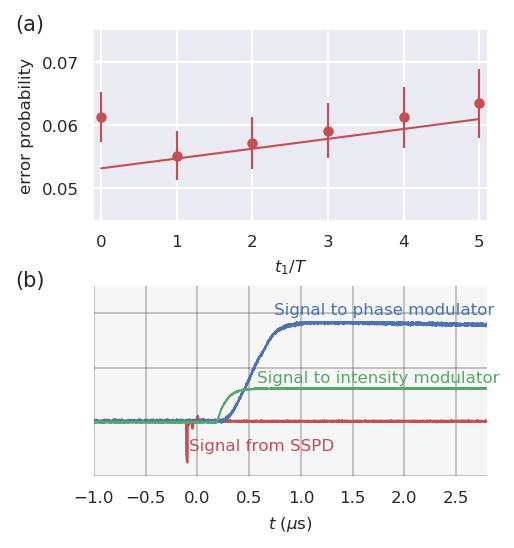}
}
\caption{
(a) Estimated error probability for the experimental result (red point) and the theoretical prediction (red line) as a function of discarding time $\Delta t$.
(b) Electric signals from SSPD (red), to the phase (blue) and intensity modulator (green).
Second time bin starts at $t=0$.
\label{Dolinar_M=2_delay}
}
\end{figure}
%%%%%%%%%%%%%%%%%%%%%%%%%%%%%%%%%%%%%%%%%%%%%%%%%%%%%%%%%%%%%%%%%%%%%%
%%%%%%%%%%%%%%%%%%%%%%%%%%%%%%%%%%%%%%%%%%%%%%%%%%%%%%%%%%%%%%%%%%%%%%
The finite time required to implement the feedback operation is a critical imperfection in practice.
While the intensity and phase modulators settle on their updated values, the displacement amplitude is not correct, thereby causing a possibility of erroneous detection outcomes.
We compensate the degradation of the performance because of the feedback delay by discarding counts observed in a time interval $\Delta t$ between each time bin. 
For the case of the single feedback operation $M=2$, we set the time width of the probe state to be $T$ = 100 $\mu$s. 
Figure \ref{Dolinar_M=2_delay}(a) depicts the error probability for $M=2$ with the first time length $t_1/T=0.31$ and the displacement magnitudes
$\{\abs{\beta_1},\abs{\beta_2}\}= \{0.71, 0.49\}$. 
The discarding time is taken into account as linear loss for the theoretical analysis, $\Delta t/T=1-\eta$, and the effect of the feedback delay is not considered because of the complexity of its analysis. 
The outcome distributions for different $\Delta t$ are obtained simultaneously and the experimentally obtained error probabilities (red points) are evaluated from the same experimental procedure.
The gap between the theoretical prediction (red line) and the experimental result for $\Delta t=0$ would be due to the delay of the feedback operation.
The error probability can be improved by discarding the counts detected in $\Delta t$ if $\Delta t \leq 1$ and degrades as $\Delta t$ further increases because the loss becomes dominant.
The electrical signals applied to the modulators are shown in Fig.~\ref{Dolinar_M=2_delay}(b).
The delay of the signal for the phase modulator can be estimated to be 1 $\mu$s, which can explain the improvement of the error probability by discarding the counts right after the first time bin.
The delay is mostly due to the setting time of the digital to analog converter employed for our experiment. 
In our experiment, the time interval $\Delta t$ is set to 2 $\mu$s which corresponds to $2\%$ of the total time length of the probe state.
A degradation of the error probability because of the $2\%$ loss is smaller than the improvement thanks to the feedback operation.

\begin{figure}[t]
\centering
{
\includegraphics[width=1.0\linewidth]
{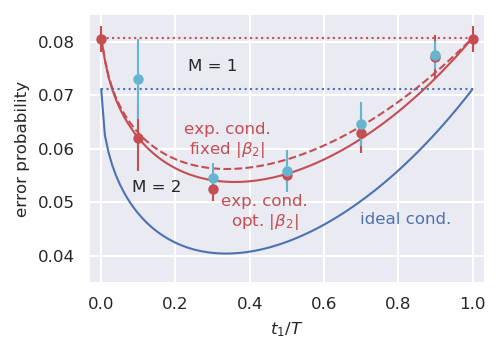}
}
\caption{
Error probability for the discrimination of the $\ket{\pm}$ states using our displacement and SPC with a single feedback operation ($M=2$) as a function of the relative temporal width of the first time bin, $t_1/T$.
\label{Dolinar_M=2_error_delay}}
\end{figure}

We show the error probability for $M=2$ as a function of the relative temporal width of the first time bin  $t_1/T$ in Fig.~\ref{Dolinar_M=2_error_delay}.
Green and red data points are obtained from the distribution of the outcomes including the counts observed in the time interval $\Delta t$ and neglecting the counts respectively.
By discarding the counts in $\Delta t$, we observe a well agreement between the experimental results
and a theoretical prediction that takes into account the visibility, the dark count and the $2\%$ loss due to the delay compensation (red solid line). 
The blue solid line represents the theoretical prediction for the ideal condition. 

%%%%%%%%%%%%%%%%%%%%%%%%%%%%%%%%%%%%%%%%%%%%%%%%%%%%%%%%%%%%%%%%%%%%%%
%%%%%%%%%%%%%%%%%%%%%%%%%%%%%%%%%%%%%%%%%%%%%%%%%%%%%%%%%%%%%%%%%%%%%%
%%%%%%%%%%%%%%%%%%%%%%%%%%%%%%%%%%%%%%%%%%%%%%%%%%%%%%%%%%%%%%%%%%%%%%
%%%%%%%%%%%%%%%%%%%%%%%%%%%%%%%%%%%%%%%%%%%%%%%%%%%%%%%%%%%%%%%%%%%%%%

For larger $M$, the degradation becomes comparable with the gain obtained from the added feedback operations, so the error probability could be saturated.
Therefore, we change the total time length as $T=100\times (M-1)\mu$s and the voltage  applied to the SSPD in order to achieve a similar dark count noise level.
The dark count rate $\nu_M$ for each experimental condition is measured to be $\{\nu_1,\nu_2,\nu_3,\nu_4,\nu_5\}=\{2.38,2.38,2.04,1.91,2.00\} \times10^{-3}$ counts/state.
Though the dark count noise is not exactly constant, 
the dependence of the error probability on the dark count noise is enough small compared with the improvement owing to the feedback operation.
A detection efficiency of the SSPD $\eta_M$ also varies depending on the voltage condition and is measured to be,
$\{\eta_1,\eta_2,\eta_3,\eta_4,\eta_5\}=\{51.4,51.4,41.1,30.7,23.6\} \%$. 

%%%%%%%%%%%%%%%%%%%%%%%%%%%%%%%%%%%%%%%%%%%%%%%%%%%%%%%%%%%%%%%%%%%%%%%%%%%%%%%%%%%%%%%%%%%%%%%%%%%%%%%%%%%%%%%%%%%%%%%%%%%%%%%%%%%%%%%%%%%%%%%%%%%%%%%%%%%%%%%%%%%%%%%%%%%%%%%%%%%%%%%%%%%%%%%%%%%%%%%%%
\section{Experimental condition of displacement magnitude}\label{Sect:3}
\begin{figure}[b]
\centering
{
\includegraphics[width=1.0\linewidth]
{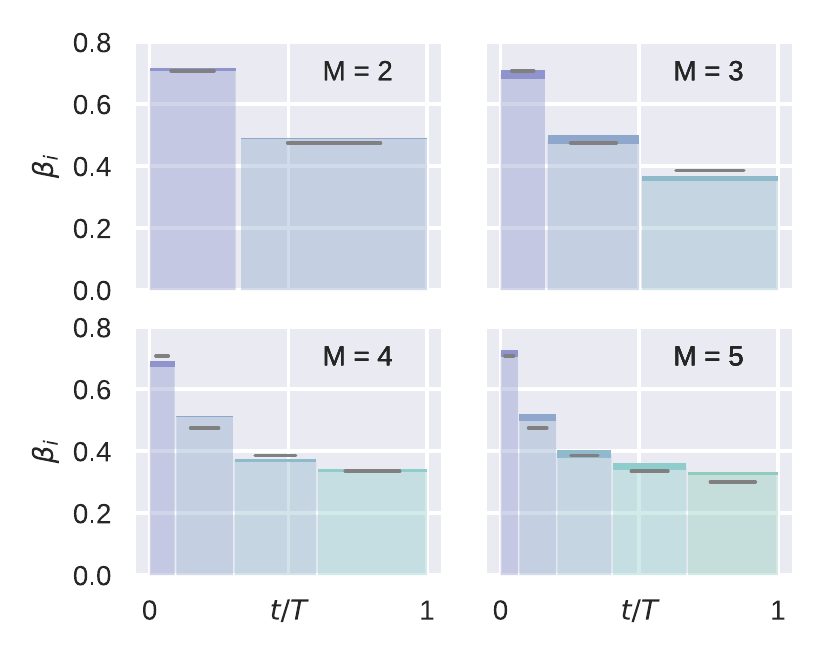}
}
\caption{
Displacement magnitude condition for Fig.4(a) in the main text.
The magnitude is set to a fixed amount at each state and the feedback timing is optimized to minimize the error probability.
\label{Dolinar_M=5_discond}
}
\end{figure}

%%%%%%%%%%%%%%%%%%%%%%%%%%%%%%%%%%%%%%%%%%%%%%%%%%%%%%%%%%%%%%%%%%%%%%%%%%%%%%%%%%%%%%%%%%%%%%%%%%%%%%%%%%%%%%%%%%%%%%%%%%%%%%%%%%%%%%%%%%%%%%%%%%%%%%%%%%%%%%%%%%%%%%%%%%%%%%%%%%%%%%%%%%%%%%%%%%%%%%%%%%
In Fig.~\ref{Dolinar_M=5_discond}, we show the displacement magnitude conditions realized in the experiment to obtain the results in Fig.4(a) in the main text.
The displacement operation at each stage is set to a fixed magnitude for technical simplicity since the gain resulting from the counting history is small.
The thickness of the top lines indicate the uncertainty ($\pm$ 1 standard deviation) of the magnitudes, while the solid gray lines show the ideal values for minimum error discrimination. 
The feedback timing is optimized to minimize the discrimination error and the time width should be shorter for earlier time bin.
The gaps between time bins are the periods that are discounted to alleviate errors due to the finite feedback time.
%%%%%%%%%%%%%%%%%%%%%%%%%%%%%%%%%%%%%%%%%%%%%%%%%%%%%%%%%%%%%%%%%%%%%%%%%%%%
%%%%%%%%%%%%%%%%%%%%%%%%%%%%%%%%%%%%%%%%%%%%%%%%%%%%%%%%%%%%%%%%%%%%%%%%%%%%%

%%%%%%%%%%%%%%%%%%%%%%%%%%%%%%%%%%%%%%%%%%%%%%%%%%%%%%%%%%%%%%%%%%%%%%%%%%%%%%%%%%%%%%%%%%%%%%%%%%%%%%%%%%%%%%%%%%%%%%%%%%%%%%%%%%%%%%%%%%%%%%%%%%%%%%%%%%%%%%%%%%%%%%%%%%%%%%%%%%%%%%%%%%%%%%%%%%%%%%%%%
\section{Maximum likelihood reconstruction of measurement operators}\label{Sect:4}
%%%%%%%%%%%%%%%%%%%%%%%%%%%%%%%%%%%%%%%%%%%%%%%%%%%%%%%%%%%%%%%%%%%%%%%%%%%%%%%%%%%%%%%%%%%%%%%%%%%%%%%%%%%%%%%%%%%%%%%%%%%%%%%%%%%%%%%%%%%%%%%%%%%%%%%%%%%%%%%%%%%%%%%%%%%%%%%%%%%%%%%%%%%%%%%%%%%%%%%%%%
%%%%%%%%%%%%%%%%%%%%%%%%%%%%%%%%%%%%%%%%%%%%%%%%%%%%%%%%%%%%%%%%%%%%%%%%%%%%
\begin{figure}[t]
\centering
{
\includegraphics[width=1.0\linewidth]
{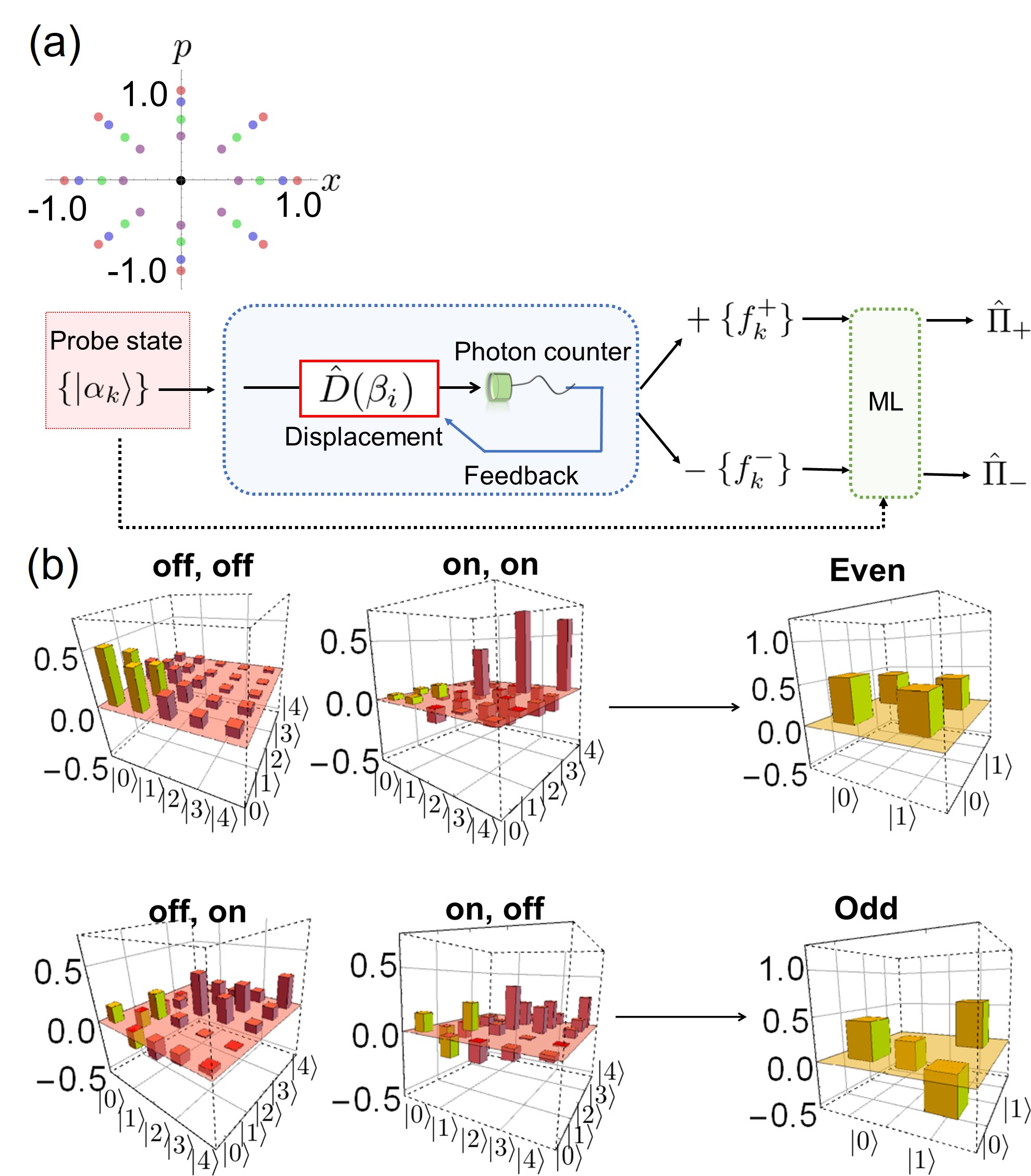}
}
\caption{
(a) Schematic of the tomography of the displacement operation with feedback operations.
A maximum likelihood reconstruction (ML) is performed using the outcome statistics $\{f_k^+,f_k^- \}$ and the characteristics of the probe states, illustrated in the phase space diagram on the left. 
(b) Experimentally reconstructed POVM elements for $M=2$. 
\label{Tomo_schematic_POVM_reconst}
}
\end{figure}
%%%%%%%%%%%%%%%%%%%%%%%%%%%%%%%%%%%%%%%%%%%%%%%%%%%%%%%%%%%%%%%%%%%%%%%%%%%%%

The most likely POVMs for the experimentally obtained data set can be estimated via a maximum likelihood reconstruction method (ML) \cite{Fiurasek}.
The log-likelihood functional defined as,
%%%%%%%%%%%%%%%%%%%%%%%%%%%%%%%%%%%%%%%%%%%%%%%%%%%%%%%%%%%%%%%%%%%%%%%%%%%%%%%
\begin{equation}
{\mathcal L}[\{\hat{\Pi}_l \}]=\sum_{l=1}^{L} \sum_{k=1}^{K} f_{kl} \ln{\tr{[\hat{\rho}_k \hat{\Pi}_l]}},
\label{eq7}
\end{equation}
%%%%%%%%%%%%%%%%%%%%%%%%%%%%%%%%%%%%%%%%%%%%%%%%%%%%%%%%%%%%%%%%%%%%%%%%%%%%%%
is maximized for the most likely POVMs, 
where $L, K, f_{kl}$ indicate the number of POVMs to be estimated, the number of probe states used for the tomography and
the experimentally obtained frequency of the outcome $l$ for the state $\hat{\rho}_k$.
The POVMs maximizing Eq.~(\ref{eq7}), under the constraints for the POVMs to be physically valid $\{ \hat{\Pi}_l\geq 0, \sum_{l=1}^{L} \hat{\Pi}_l =\hat{I} \}$, can be found by recursively applying the following transformation,
%%%%%%%%%%%%%%%%%%%%%%%%%%%
\begin{equation}
\hat{\Pi}_l= \hat{\lambda}^{-1} \hat{R}_l \hat{\Pi}_l \hat{R}_l \hat{\lambda}^{-1},
\label{eq8}
\end{equation}
%%%%%%%%%%%%%%%%%%%%%%%%%%%
where
%%%%%%%%%%%%%%%%%%%%%%%%%%%
\begin{eqnarray}
\hat{R}_l&=& \sum_{k=1}^{K} \frac{f_{kl}}{\tr{[\hat{\rho}_k \hat{\Pi}_l]}}\hat{\rho}_k, 
\\
\hat{\lambda}&=& (\sum_{l=1}^L \hat{R}_l \hat{\Pi}_l \hat{R}_l)^{1/2}.
\label{eq9}
\end{eqnarray}
%%%%%%%%%%%%%%%%%%%%%%%%%%%

A simple schematic of the tomography is depicted in Fig.~\ref{Tomo_schematic_POVM_reconst}(a).
We prepare 33 different probe coherent states ($K=33$) with 4 weak amplitude conditions $\abs{\alpha}\approx\ \{0.4, 0.6, 0.8, 1.0 \}$ and 8 phase conditions $j\pi/4,\ j=0,\ldots,7$ in addition to the vacuum state.
Experimentally reconstructed POVM elements for $M=2$ with the displacement amplitudes and the feedback timing $\{\beta_1,\beta_2^{\mathrm{off}},\beta_2^{\mathrm{on}}\}= \{-0.63, -0.51, 0.39\}$ and $t_1/T=0.30$ are shown in Fig.~\ref{Tomo_schematic_POVM_reconst}(b).
They are first reconstructed in 5-dimensional space in the photon number basis and then truncated to two dimensions to evaluate the discrimination error.
Our measurement provides $2^M$ outcomes for a given number of $M$ and thee error probability is finally calculated using the POVMs for the parity of the total number of ``on'' events.